\documentclass[preprint,aps,nofootinbib]{revtex4}
\usepackage{graphicx}
\usepackage{amsmath,amssymb}
\usepackage{longtable}
\begin{document}
\draft
\title{Long-range entanglement in the XXZ Heisenberg spin chain after a local quench}
\date{\today}
\author{Jie Ren$^{1}$\footnote{E-mail: jren@cslg.edu.cn}}
\author{Shiqun Zhu$^2$\footnote{E-mail: szhu@suda.edu.cn}}

\affiliation{$^1$Department of Physics and Jiangsu Laboratory of
Advanced Functional Materials, Changshu Institute of
Technology, Changshu, Jiangsu 215500, China\\
$^2$School of Physical Science and Technology, Suzhou University,
Suzhou, Jiangsu 215006, China}

\begin{abstract}

The long-range entanglement dynamics of an one-dimensional spin-1/2
anisotropic XXZ model are studied using the method of the adaptive
time-dependent density-matrix renormalization-group. The long-range
entanglement can be generated when a local quench on one of boundary
bonds is performed in the system. The anisotropic interaction has a
strong influence both on the maximal value of long-range
entanglement and the time of reaching the maximum long-range
entanglement. The local coupling has a notable impact on the
long-range entanglement, but it can be neglected in the time of
reaching the maximal long-range entanglement.

\vskip 0.6 cm


\vskip 0.6 cm

\end{abstract}

\maketitle

\section{Introduction}

Entanglement generation and distribution is one of the important
problems in performing quantum-information tasks, such as quantum
computation and quantum teleportation
\cite{Nielsen,Bennett,Wootters}. Many results showed that
entanglement existed naturally in the spin chain when the
temperature is at zero \cite{Osborne,Osterloh}. It is discouraging
that the entanglement in many spin systems is typically very short
ranged. It exists only in the nearest neighbors and the next nearest
neighbors \cite{Amico}. It is interesting that some schemes for
long-range entanglement, such as exploiting weak couplings of two
distant spins to a spin chain were proposed
\cite{Venuti01,Venuti02,Giampaolo}. These methods have limited
thermal stability or very long time scale of entanglement
generation. In recent years, many researches has shifted the focus
of the study in the dynamics of entanglement
\cite{Bose0,Hartmann,Li,A,Bose01}. The dynamics of long range
entanglement are obtained by a time quench of magnetic field
\cite{Galve,Wang}. The end-to-end entanglement can be shared across
a chain of arbitrary range \cite{Bose0}. It is excited that the most
common situation studies so far concerns a sudden quench of some
coupling of the model Hamiltonian \cite{Chiara,Bose02,Bose03}. In
Ref. \cite{Bose03}, it showed that the long-range entanglement can
be engineered by a non-perturbative quenching of a single bond in a
Kondo spin chain with impurity. This is the first example that a
minimal local action on a spin chain can generate long-range
entanglement dynamically. It would be interesting to investigate the
possibility of producing long-range entanglement in the XXZ
Heisenberg spin chain without impurity.

It is well-known that the Hamiltonian of an opened chain of N
spin-$1/2$ system with nearest-neighbor XXZ interaction is given by

\begin{equation}
\label{eq1}H_0=\sum_{i=1}^{N-1}J[S^x_{i}S^x_{i+1}
+S^y_{i}S^y_{i+1}+\Delta S^z_{i}S^z_{i+1}],\\
\end{equation}
where $S^{\alpha}_i(\alpha=x, y, z)$ are spin operators on the
$i$-th site, $N$ is the length of the spin chain. The parameter $
\Delta$ denotes the couplings in the $z$-axis.This model can be
realized in the Josephson-junction\cite{Fazio}and optical
lattices\cite{Duan,Trotzky}.

In the paper, the long-range entanglement in the XXZ Heisenberg spin
chain after a local quench is investigated. An opened boundary
condition (OBC) is assumed because the antiferromagnetic Heisenberg
spin chain with OBC can be achieved artificially in the experiment
\cite{hirjibehedin}, and the coupling $J=1$ is considered for
simplicity. In section II, the local quench is presented. The
concurrence is used as a measurement of the entanglement. In section
III, a more general situation and a single quench and its effect on
the end-to-end qubits entanglement are analyzed. The effects of the
anisotropic interaction and system size on the end-to-end qubits
entanglement is also studied. The robustness of the entanglement
against an increase in temperature is investigated. A discussion
concludes the paper.

\section{Local Quench and Entanglement Measure}

In the paper, the system is assumed to be in the ground state $Gs_0$
of $H_0$ initially. A local quench change of the coupling between
the first qubit and the second qubit with the same anisotropy
interaction $\Delta$ in Eq. (1) is performed. The Hamiltonian of the
system modifies to

\begin{equation}
\label{eq2}H_1=J_1[S^x_{1}S^x_{2} +S^y_{1}S^y_{2}+\Delta
S^z_{1}S^z_{2}]+\sum_{i=2}^{N-1}J[S^x_{i}S^x_{i+1}
+S^y_{i}S^y_{i+1}+\Delta S^z_{i}S^z_{i+1}].\\
\end{equation}
Since $[H_0,H_1]\neq 0$, the the ground state $Gs_0$ of $H_0$ is not
one of the eigenstates of $H_1$. The state of the system will evolve
as
\begin{equation}
\label{eq3}\psi(t)=\exp^{-iH_1t}Gs_0.
\end{equation}

In the paper, the concurrence is chosen as a measurement of the
pairwise entanglement \cite{Wootters}. The concurrence $C$ is
defined as

\begin{equation}
\label{eq4} C_{1,N} = \max \{{\lambda_1 - \lambda_2 - \lambda_3 -
\lambda_4 ,0}\},
\end{equation}
where the quantities $\lambda_i (i=1, 2, 3, 4)$ are the square roots
of the eigenvalues of the operator $\varrho = \rho_{12}(\sigma_1^y
\otimes \sigma_N^y)\rho_{1,N}^\ast (\sigma_1^y \otimes \sigma_N^y)$.
They are in descending order. The case of $C_{1,N}=1$ corresponds to
the maximum entanglement between the two qubits, while $C_{1,N}=0$
means that there is no entanglement between the two qubits.

\section{Long-Range Entanglement Dynamics}

It is known that it is hard to calculate the dynamics of
entanglement because of the lack of knowledge of eigenvalues and
eigenvectors of the Hamiltonian. For models that are not exactly
solvable, most of researchers resort to exact diagonalization to
obtain the ground state for small system size. The method is
difficult to be applied for large system size $N > 20$. For a large
system, the adaptive time-dependent density-matrix
renormalization-group can be applied with a second order Trotter
expansion of the Hamiltonian as described in \cite{White,Vidal02}.
In order to check the accuracy of the results of the adaptive
time-dependent density-matrix renormalization-group, the results of
exact diagonalization can be considered as a benchmark for a small
size system.

The numerical error of the adaptive time-dependent density-matrix
renormalization-group comes from the discarded weight and the
Trotter decomposition. The error of discarded weight is dependent on
the data precision and truncated Hilbert space. The error of the
Trotter decomposition is relied on Trotter slicing. In our numerical
simulations a Trotter slicing $\delta t=0.05$ and Matlab codes of
the adaptive time-dependent density-matrix renormalization-group
with double precision are performed with a truncated Hilbert space
of $m=100$. In turns out that a typically discarded weight of
$\delta \rho \leq 10^{-8}$. The error of Trotter decomposition
$\delta \propto (\delta t)^3$. These can keep the relative error
$\delta C$ in $C$ below $10^{-3}$ for a chain of $N= 60$ sites with
time $t\leq50/J$.

In the adaptive time-dependent density-matrix renormalization-group,
it is hard to calculate the reduced density matrix $\rho_{1,N}$. By
making use of the relation between the correlation and the reduced
density matrix, the reduced density matrix can be expressed as

\begin{equation}
\label{eq5} \rho_{1,N}=\frac{1}{4}
[I_{1,N}+\sum_{i,j=x,y,z}\langle\sigma_1^i\sigma_{N}^j\rangle\sigma_1^i\sigma_N^j],
\end{equation}
where $\sigma^{\alpha}_k(\alpha=x, y, z)$ are Pauli operators and
$I_{1,N}$ is identity matrix.

We obtain the reduced density matrix $\rho_{1,N}$ by Eq. (5), then
calculate the entanglement between two-ends qubits. The end-to-end
entanglement $C_{1, N}$ is plotted as a function of the anisotropic
interaction $\Delta$ and time $t$ with $J_1=-0.01$ in Fig. 1(a). It
is seen that the long-range entanglement $C_{1, N}$ can be generated
after a short time. When the time $t$ increases, the entanglement
reaches the maximal value and then disappears quickly. That is,
there is a peak in the entanglement. The peak of the end-to-end
entanglement is plotted as a function of the anisotropic interaction
$\Delta$ in Fig. 1(b). It is seen that the peak in long-range
end-to-end entanglement increases with the anisotropic interaction
$\Delta$. It reaches the maximal value when the anisotropic
interaction $\Delta=1$ \cite{Gu,Bose04}. With the anisotropic
interaction $\Delta$ increases further, the height of the peak
decreases.

The relation between the time of reaching the maximal end-to-end
entanglement labeled by $T_{max}$ and the anisotropic interaction
$\Delta$ can be seen in the inset of Fig. 1(b). The time of reaching
the maximal long-range entanglement decreases when the anisotropic
interaction $\Delta$ increases, while they are also not a linear
relationship. Similar to Refs. \cite{Bose0,Bose03}, the system
generates end-to-end entanglement periodically. In our simulations,
the error of the simulations increases when the time increases.
Since the coherent time of the system is not very long, the
long-distance entanglement is not shown when the time $t>50/J$. It
is shown that a sharp drop of long-distance entanglement occurs at
$\Delta=-0.5$. Furthermore, the time of the long-range entanglement
reaching the maximal value goes up drastically. This is similar to
the anomalous behavior appeared in Ref. \cite{Bose04}.

The end-to-end entanglement $C_{1, N}$ is plotted as a function of
the interaction $J_1$ and the time $t$ with $\Delta=1$ in Fig. (2).
The size of the system $N=20$ is chosen. It does not include the
case of $J_1=0$. The long-range entanglement $C_{1, N}$ can be
generated after a short time. There is a peak in $C_{1,N}$ when
$C_{1,N}$ is plotted as a function of time $t$. It is seen that the
influence of the coupling $J_1$ on the maximal long-range
entanglement is relatively large. The peak of the end-to-end
entanglement increases when the coupling interaction $J_1$
increases, and reaches the maximal value when $J_1=-0.1$. It seems
that the changing of local interaction from antiferromagnetic to
ferromagnetic may enhance the entanglement creation \cite{Venuti01}.
When $J_1$ increases further to $J_1>0.32$, the long-range
entanglement disappears. It is interesting that the coupling
interaction $J_1$ has a relatively small impact on the time when the
end-to-end entanglement generates and reaches the maximal value.

It is easy to obtain that $[H_0, \sum_{i=1}^N S^z_i] =[H_1,
\sum_{i=1}^N S^z_i] = 0$. It means that the ground states of $H_0$
is a total singlet $S_{tot}=0$. The Hamiltonian $H_1$ has invariant
on every excitation subspaces. During the evolution, the boundary
spin at $i=2$ will have a strong tendency to form a singlet pair
with its only nearest neighbor $i=3$ on the right-hand side. This is
similar for spin pairs $(4,5)$ and $(6,7)$, ......, etc. Then the
two-end spin qubits form a singlet. The local interaction $J_1$
decides the ability of forming singlet for even bands. Thus, the
long-distance entanglement creates.

The thermalization and relaxation during the period of generating
long-distance entanglement are neglected because the dynamical time
scale is quite short. When initial state is taken to be the relevant
thermal state, the end-to-end entanglement is plotted as a function
of temperature in Fig. 3 after bond quenching for different system
sizes. It is shown that the entanglement vanishes when $kT>1.16$
with the size of $N=8$, and $kT>1.06$ with the size of $N=10$. It
seems that our scheme is quite robust against temperature and is
similar to that in Ref. \cite{Bose03}. The initial thermal state
does not have long-distance entanglement in Eq. (1), but the energy
gap between the ground state and the first and other excited states
are larger than the system with impurity. This leads to our scheme
is quite robust against temperature.

It is interesting to investigate the long-range entanglement
creation even for very long chain of large size. The maximal value
of the end-to-end entanglement $C_{1, N}$ is plotted as a function
of the size of the system for different anisotropic interaction
$\Delta$ and different interaction $J_1$ in Fig. 4(a). It is found
that the maximal value of the end-to-end entanglement $C_{1, N}$
decreases when the size of system increases. When the anisotropic
interaction $\Delta=1$ and the interaction $J_1=-0.1$, the maximal
value is 0.5481 for $N=40$ and 0.4547 for $N=60$. It is shown that
the maximal value of the end-to-end entanglement label by $\xi(N)$
is given by \cite{Bose0}

\begin{equation}
\label{eq6}\xi(N)\simeq1.35N^{-1/3}.
\end{equation}
In Ref. \cite{Bose0}, $\xi(40)\simeq0.3947$ for $N=40$, and
$\xi(40)\simeq0.3448$ for $N=60$. While in the Kondo regime,
$\xi(40)$ can be as large as $0.7$ for $N=40$. It seems that our
results are also quite good. The long distance entanglement can be
obtained by performing a local quench in either one of two boundary
bands. The time of reaching the maximal end-to-end entanglement
labeled by $T_{max}$ is plotted as a function of the size $N$ of the
system for different anisotropic interaction $\Delta$ and different
interaction $J_1$ in Fig. 4(b). The time of the end-to-end
entanglement reaching the maximal value $T_{max}$ is linear increase
when the size $N$ increases \cite{Bose0,Bose04}. It is seen that the
slope of the line is almost dependent on the anisotropic interaction
$\Delta$.

In the paper, adiabatic quenches are studied and the decoherence
effects in the system are ignored. It is noted that the XXZ spin
chains can be realized in the Josephson-junction array
\cite{Fazio}and optical lattices \cite{Duan,Trotzky}. In
Josephson-junction array, the interactions and the anisotropy can be
modulated by varying voltages \cite{Allcock}. The Josephson-junction
array system does not have significant decoherence over our
time-scales ($T_{max} \simeq N/J$) \cite{Lyakhov}. The anisotropic
interaction quenches can be achieved successfully with long
decoherence time in the optical lattices experiment \cite{Trotzky}.

\section{Discussion}

By using the method of the adaptive time-dependent density-matrix
renormalization-group, the time evolution of the entanglement in a
one-dimensional spin-1/2 anisotropic XXZ model is investigated when
a local quench is performed in the system. The local quench is a
abrupt change of interaction between the first qubit and the second
qubit. The dynamics of pairwise entanglement between the two ends
qubits in the spin chain is studied. The entanglement of the
two-ends spin qubits can be created  after the local quench is
performed.The time when the long-range entanglement generates
decreases in a nonlinear manner with the anisotropic interaction
 increases. It reaches the maximal value when the
anisotropic interaction $\Delta=1$. The maximal value of the
end-to-end entanglement increases  with the coupling interaction
$J_1$ increases, and reaches the maximal when $J_1=-0.1$. It is
interesting that the coupling interaction has a relatively small
impact on the time when the long-range entanglement reaches the
maximal value. This phenomenon may be used to control the dynamics
of the entanglement by varying the anisotropic interaction and the
local quench interaction of the Heisenberg spin chain.

\vskip 0.4 cm

{\textbf{Acknowledgments}}

It is a pleasure to thank Yinsheng Ling and Yinzhong Wu for their
many helpful discussions. The financial support from the National
Natural Science Foundation of China (Grant No. 10774108) is
gratefully acknowledged.

\clearpage
\newpage
\begin{figure}
\includegraphics[scale=0.7]{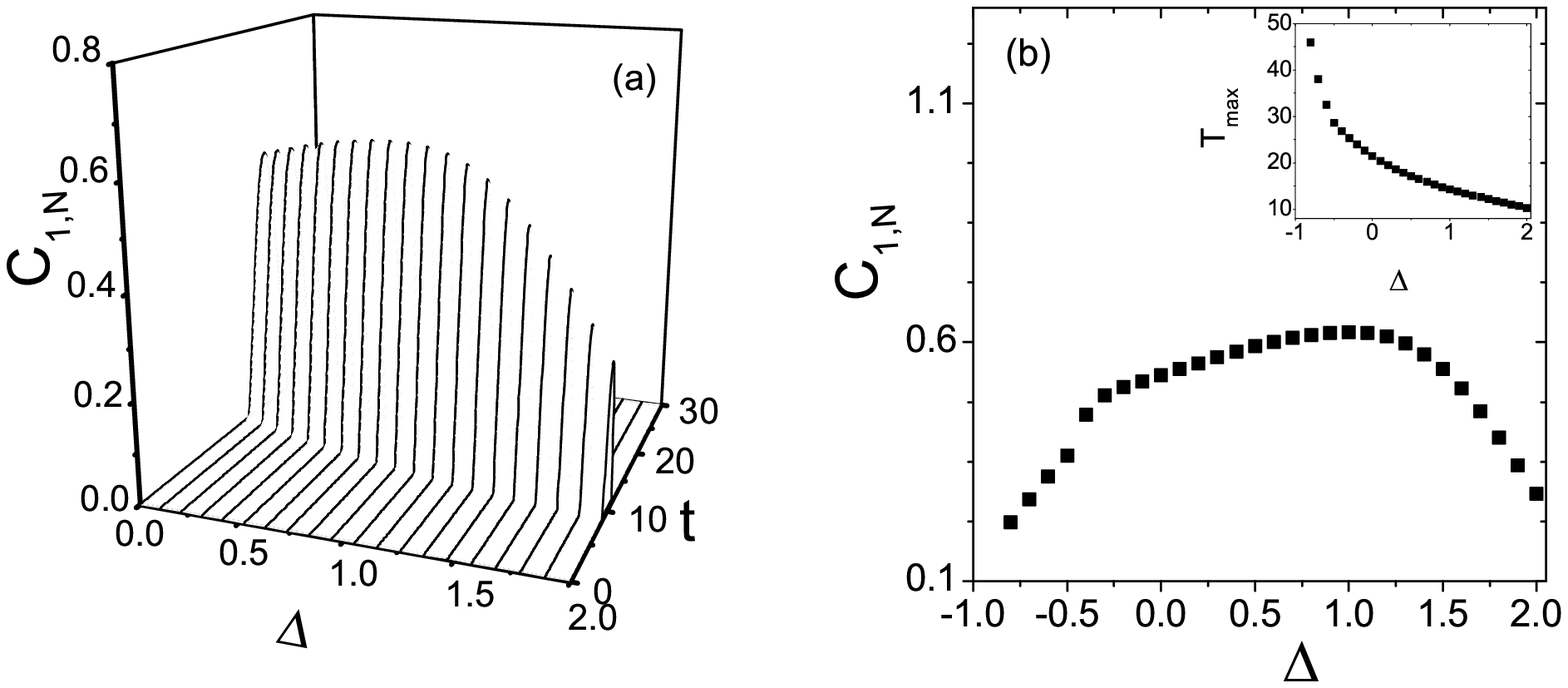}\caption{(a). The long-range
entanglement $C_{1, N}$ is plotted as a function of anisotropic
interaction $\Delta$ and time $t$ with $J_1=-0.1$. The size of the
system is $N=20$. (b). The maximal value of the end-to-end
entanglement is plotted as a function of the anisotropic interaction
$\Delta$. The inset shows the time of reaching the maximal
end-to-end entanglement labeled by $T_{max}$ as a function of
anisotropic interaction $\Delta$.}
\end{figure}

\clearpage
\newpage
\begin{figure}
\includegraphics[scale=0.7]{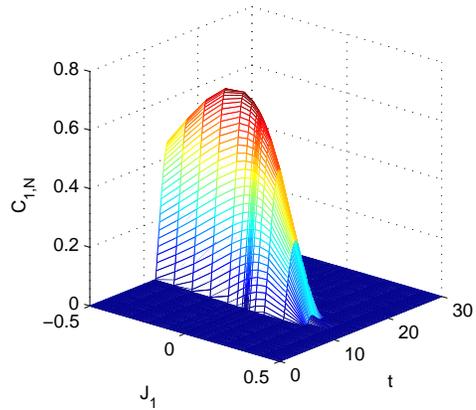}\caption{The long-range entanglement
$C_{1, N}$ is plotted as a function of interaction $J_1$ and time
$t$ when there is local quench of the interaction $J_1$. The size of
the system is $N=20$ and the anisotropic interaction $\Delta=1$. It
does not include the case of $J_1=0$. }
\end{figure}

\clearpage
\newpage
\begin{figure}
\includegraphics[scale=0.85]{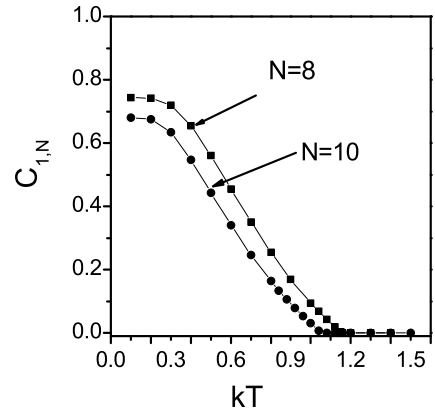}\caption{The maximal value of end-to-end entanglement is plotted as a function of
temperature after band quenching for different sizes.}
\end{figure}

\clearpage
\newpage
\begin{figure}
\includegraphics[scale=0.7]{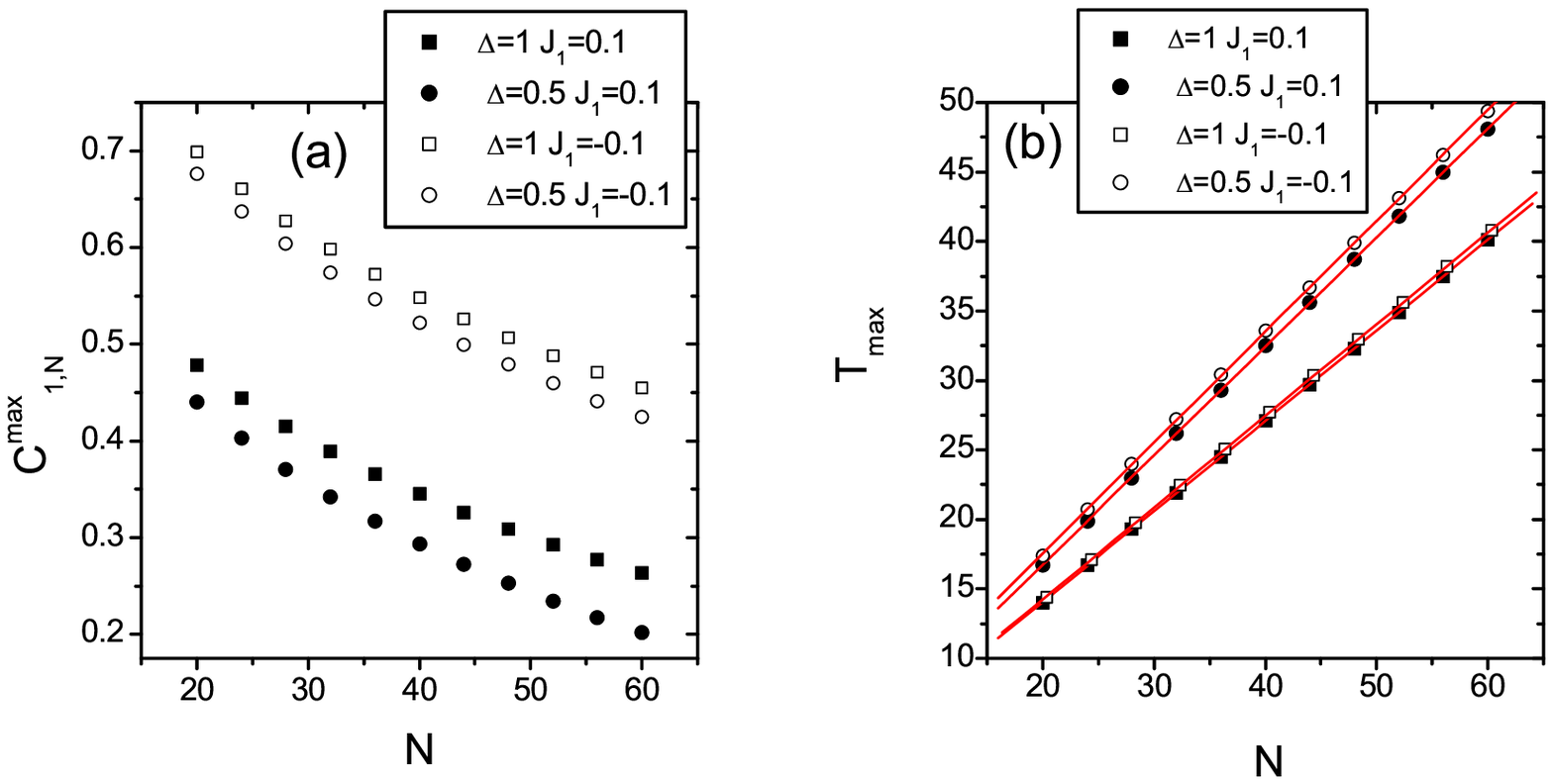}\caption{ (a). The maximal value of the long-range entanglement $C_{1, N}$ is
plotted as a function of the size of the system for different
anisotropic interaction $\Delta$ and different interaction $J_1$.
(b). The time of reaching the maximal long-range entanglement
labeled by $T_{max}$ is plotted as a function of the size of the
system for different anisotropic interaction $\Delta$ and different
interaction $J_1$. The red lines are fixed lines.}
\end{figure}


\begin{references}

\bibitem{Nielsen} M. A. Nielsen and I. L. Chuang, Quantum Computation and Quantum
Information (Cambridge University Press, Cambridge, England, 2000).

\bibitem{Bennett} C. H. Bennett, G. Brassard, C. Crepeau, R. Jozsa, A. Peres, and
W. K. Wootters, Phys. Rev. Lett. \textbf{70}, 1895 (1993).

\bibitem{Wootters} W. K. Wootters, Phys. Rev. Lett. \textbf{80}, 2245 (1998).

\bibitem{Osborne} T. J. Osborne and M. A. Nielsen, Phys. Rev. A
\textbf{66}, 032110 (2002).

\bibitem{Osterloh} A. Osterloh, L. Amico, G. Falci, and R. Fazio,
Nature \textbf{416}, 608 (2002).

\bibitem{Amico} L. Amico, R. Fazio, A. Osterloh, and V. Vedral, Rev. Mod. Phys. \textbf{80}, 517 (2008).

\bibitem{Venuti01}L. C. Venuti, C. D. E. Boschi, and M. Roncaglia, Phys. Rev. Lett. \textbf{96}, 247206
(2006).

\bibitem{Venuti02}L. C. Venuti, S. M. Giampaolo, F. Illuminati, and P. Zanardi, Phys. Rev. A \textbf{76}, 052328 (2007).

\bibitem{Giampaolo} S. M. Giampaolo and F. Illuminati, New J. Phys. \textbf{12}, 025019
(2010).

\bibitem{Bose0} S. Bose, Phys. Rev. Lett. \textbf{91}, 207901
(2003).

\bibitem{Hartmann} M. J. Hartmann, M. E. Reuter, and M. B. Plenio, New J. Phys. \textbf{8}, 94
(2006).

\bibitem{Li} Y. Li, T. Shi, B. Chen, Z. Song, and C.-P. Sun, Phys. Rev. A \textbf{71}, 022301
(2005).

\bibitem{A} A. W\'{o}jcik, T. {\L}uczak, P. Kurzy\'{n}ski, A. Grudka, T. Gdala, and M. Bednarska, Phys. Rev. A \textbf{72}, 034303 (2005).

\bibitem{Bose01} J. Reslen and S. Bose, Phys. Rev. A \textbf{80}, 012330
(2009).

\bibitem{Galve} F. Galve, D. Zueco, S. Kohler, E. Lutz, and P.
H\"{a}nggi, Phys. Rev. A \textbf{79}, 032332 (2009).

\bibitem{Wang} X. Wang, A. Bayat, S. G. Schirmer, S. Bose, Phys.
Rev. A \textbf{81}, 032312 (2010)

\bibitem{Chiara} G. D. Chiara, S. Montangero, P. Calabrese, and R. Fazio, J. Stat.
Mech. P03001 (2006).

\bibitem{Bose02} H. Wichterich and S. Bose, Phys. Rev. A \textbf{79}, 060302(R)
(2009).

\bibitem{Bose03} P. Sodano, A. Bayat, and S. Bose, Phys. Rev. B \textbf{81},
100412(R) (2010).

\bibitem{Fazio} R. Fazio and H. van der Zant, Phys. Rep. \textbf{355}, 235 (2001).

\bibitem{Duan} L.-M. Duan, E. Demler, and M. D. Lukin, Phys. Rev. Lett. \textbf{91}, 090402
(2003).

\bibitem{Trotzky} S. Trotzky, P. Cheinet, S. F\"{o}lling, M. Feld, U.
Schnorrberger, A. M. Rey, A. Polkovnikov, E. A. Demler, M. D. Lukin,
and I. Bloch, Science \textbf{319}, 295 (2008).

\bibitem{hirjibehedin} C. F. Hirjibedin, C. P. Lutz, and A. J. Heinrich, Science \textbf{312},
1021 (2006).

\bibitem{White} S. R. White and A. E. Feiguin, Phys. Rev. Lett. \textbf{ 93}, 076401
(2004).

\bibitem{Vidal02} G. Vidal, Phys. Rev. Lett. \textbf{93}, 040502 (2004).

\bibitem{Gu} S.-J. Gu, G.-S. Tian, and H.-Q. Lin, Phys. Rev. A \textbf{71},
052322 (2005)

\bibitem{Bose04} A. Bayat and S. Bose, Phys. Rev. A \textbf{81}, 012304
(2010).

\bibitem{Allcock} J. Allcock and N. Linden, Phys. Rev. Lett. \textbf{102}, 110501
(2009).

\bibitem{Lyakhov} A. Lyakhov and C. Bruder, New J. Phys. \textbf{7}, 181 (2005).

\end{references}
\end{document}